\documentclass[conference]{IEEEtran}
\IEEEoverridecommandlockouts

\usepackage{cite}
\usepackage{amsmath,amssymb,amsfonts}
\usepackage{algorithmic}
\usepackage{graphicx}
\usepackage{textcomp}
\usepackage{xcolor}
\usepackage{url}
\def\BibTeX{{\rm B\kern-.05em{\sc i\kern-.025em b}\kern-.08em
    T\kern-.1667em\lower.7ex\hbox{E}\kern-.125emX}}
\begin{document}

\title{A Transformer-Based Approach for Diagnosing Fault Cases in Optical Fiber Amplifiers\\
\thanks{This work is supported by the SNS Joint Undertaken under grant agreement No. 101096120 (SEASON). Responsibility for the content of this publication is with the authors.

\copyright2025 IEEE. Personal use of this material is permitted. Permission from IEEE must be obtained for all other uses, in any current or future media, including reprinting/republishing this material for advertising or promotional purposes, creating new collective works, for resale or redistribution to servers or lists, or reuse of any copyrighted component of this work in other works.}
}

\author{\IEEEauthorblockN{Dominic Schneider}
\IEEEauthorblockA{\textit{Advanced Technology} \\
\textit{Adtran Networks SE}\\
98617 Meiningen, Germany \\
dominic.schneider@adtran.com}
\and
\IEEEauthorblockN{Lutz Rapp}
\IEEEauthorblockA{\textit{Advanced Technology} \\
\textit{Adtran Networks SE}\\
98617 Meiningen, Germany \\
lutz.rapp@adtran.com}
\and
\IEEEauthorblockN{Christoph Ament}
\IEEEauthorblockA{\textit{Faculty of Applied Computer Science} \\
\textit{University of Augsburg}\\
86159 Augsburg, Germany \\
christoph.ament@uni-a.de}
}

\maketitle

\begin{abstract} 
A transformer-based deep learning approach is presented that enables the diagnosis of fault cases in optical fiber amplifiers using condition-based monitoring time series data. The model, Inverse Triple-Aspect Self-Attention Transformer (ITST), uses an encoder-decoder architecture, utilizing three feature extraction paths in the encoder, feature-engineered data for the decoder and a self-attention mechanism. The results show that ITST outperforms state-of-the-art models in terms of classification accuracy, which enables predictive maintenance for optical fiber amplifiers, reducing network downtimes and maintenance costs.
\end{abstract}


\section{Introduction} 
In present optical transmission links, optical fiber amplifiers are key components in long-haul and metro fiber optical networks. Aging of these devices can result in slowly but permanently increasing performance degradation, but also complete outage of the affected link, resulting in cost-intensive maintenance and high financial loss of income. Predictive maintenance comprises an advanced system monitoring technique that includes the early detection of anomalous behavior and the diagnosis of the cause of the fault. The application to optical fiber amplifiers enables the planning of maintenance measures before a total failure of the system occurs and thus minimizes the risk of network failures.

The methods for diagnosing fault cases (FCs) are grouped into analytical and data-driven approaches. The creation of analytical representations for modeling the influences of degrading components on the overall system involves a high level of effort as the degree of complexity of the systems to be diagnosed is very high\cite{frank2000model}. In contrast to analytical methods, data-driven methods use condition-based monitoring (CBM) data of the device. Classic machine learning (ML) algorithms such as artificial neural networks \cite{mohd2020neural} have shown a good applicability. Deep learning (DL) has an intrinsic feature extraction mechanism and scales better than shallow ML algorithms, especially with large amounts of data. In the beginning, fully connected DL structures such as deep belief networks \cite{gan2016construction} and stacked auto-encoders \cite{shao2017novel} were used, but they have a high utilization of resources. Convolutional neural networks (CNN) \cite{zhao2020intelligent} reduce the parameter space through sparse connections and pooling techniques. Analogous to CNNs, long short-term memory (LSTM) networks have developed, which enable the recording of temporal dependencies through the implementation of memory mechanisms. The combination of CNN and LSTM led to models with improved feature extraction capabilities and capturing of long-term dependencies, such as convolutional bi-directional long short-term memory (CBLSTM) networks \cite{zhao2017learning}. Yet, the feature extraction of convolutional layers is limited with respect to the distinction between local and global dependencies. Transformers use an encoder-decoder architecture and can capture local and global dependencies through the attention mechanism. As a result, models such as Diagnosisformer \cite{hou2023diagnosisformer} enable a more accurate diagnosis of FCs compared to other data-driven methods.

This work presents two major contributions. First, a novel transformer-based model, Inverse Triple-Aspect Self-Attention Transformer (ITST), is presented, which uses three feature extraction paths in the encoder. Each path extracts features from the CBM data in the time, sensor and frequency domain, respectively. The decoder uses feature-engineered data. Both, the encoder and decoder path utilize the self-attention mechanism. In combination, this improves the accuracy of the diagnosis by up to 0.05 points compared to state-of-the-art methods. Second, the diagnosis of FCs using this novel model for optical fiber amplifiers is exemplified using C-band erbium-doped fiber amplifiers (EDFAs) for the first time. This enables the early detection of degrading components, allowing maintenance measures to be planned and performed, before the system fails.

\section{Methodology} 
\begin{figure*}[ht!]
    \centering
    \includegraphics[width=0.82\linewidth]{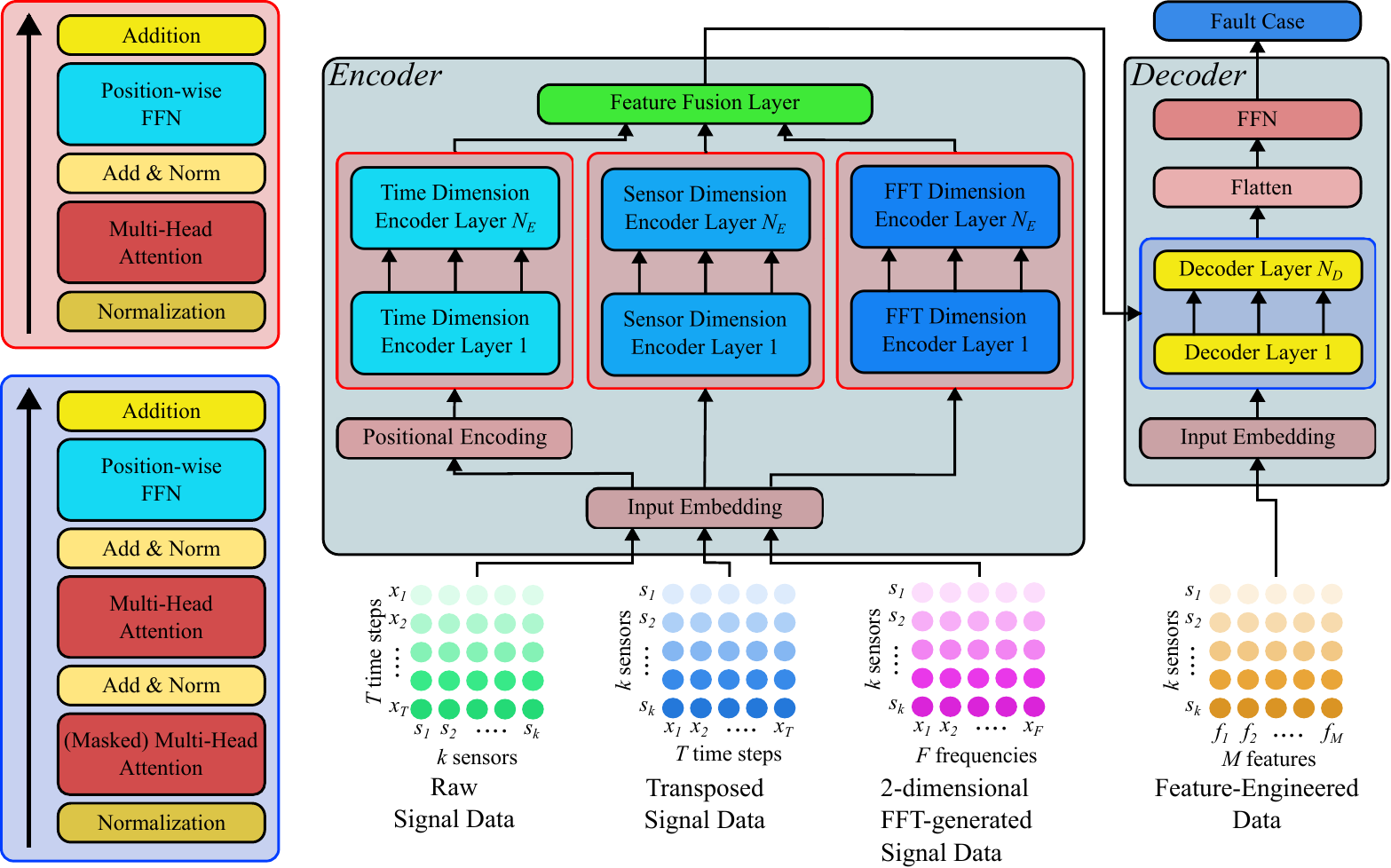}
    \caption{Deep learning model architecture of ITST}
    \label{fig:1}
\end{figure*}
The diagnosis of FCs can be understood as a classification task in a data-driven context. Here, the CBM data of the system are used at runtime. The data are available in the form $X_{t}\in\mathbb{R}^{k}$ with $t=\left(1,\dots,T\right)$, where $T$ represents the number of measurement points and $k$ is the number of parameters. The classification can be understood as a mapping of the input data $X_{t}$ to a vector $y_{t}=f\left(X_{t}\right)$, where the function $f$ represents an arbitrary model. The elements of vector $y_{t}=\left[y_{t}^{class_{1}},\dots,y_{t}^{class_{c}}\right]$ describe a probability value, how strong the input data belongs to a class $c\in C$ that indicates the FC of the system.

The encoder-decoder architecture of ITST is used as mapping function $f$, shown in Fig.~\ref{fig:1}. The encoder consists of three feature extraction paths. The first path uses the unmodified CBM data and generates a feature map in the time domain. The second path transposes the CBM data matrix and thus enables feature extraction in the sensor domain. The last path applies a two-dimensional Fast Fourier transform (FFT) to the data, generating a feature map in the frequency domain. These three feature maps are then concatenated and made available to the decoder. Each feature extraction path of the encoder utilize $N$ encoder layer, consisting of a multi-head attention and a position-wise feed forward network (FFN) with residual connections, highlighted in red on the left in Fig.~\ref{fig:1}.

The decoder uses feature-engineered data derived from the CBM data matrix along the time dimension, such as the statistical mean, the statistical variance and the three fitted parameters of a second-order polynomial regression. Each decoder layer consists of a multi-head attention, a position-wise FFN and a cross-attention mechanism with the encoder output, highlighted in blue on the left in Fig.~\ref{fig:1}. The encoder and decoder paths utilize the self-attention mechanism, which enables the model to capture local and global dependencies in the data.

Furthermore, the model is trained using the Adam optimizer and the learning rate is scheduled accordingly \cite{vaswani2017attention}. The evaluation metric is the categorical cross-entropy (CC) and the best choice for the hyperparameters are found using Bayesian optimization. The exact structure and training routine of ITST can be found at \url{https://github.com/DomSchResearch/ITST} and the trained model is available at \url{https://huggingface.co/dschneider96/ITST}

\section{Data Acquisition} 
\begin{figure}[b!]
    \centering
    \includegraphics[width=\linewidth]{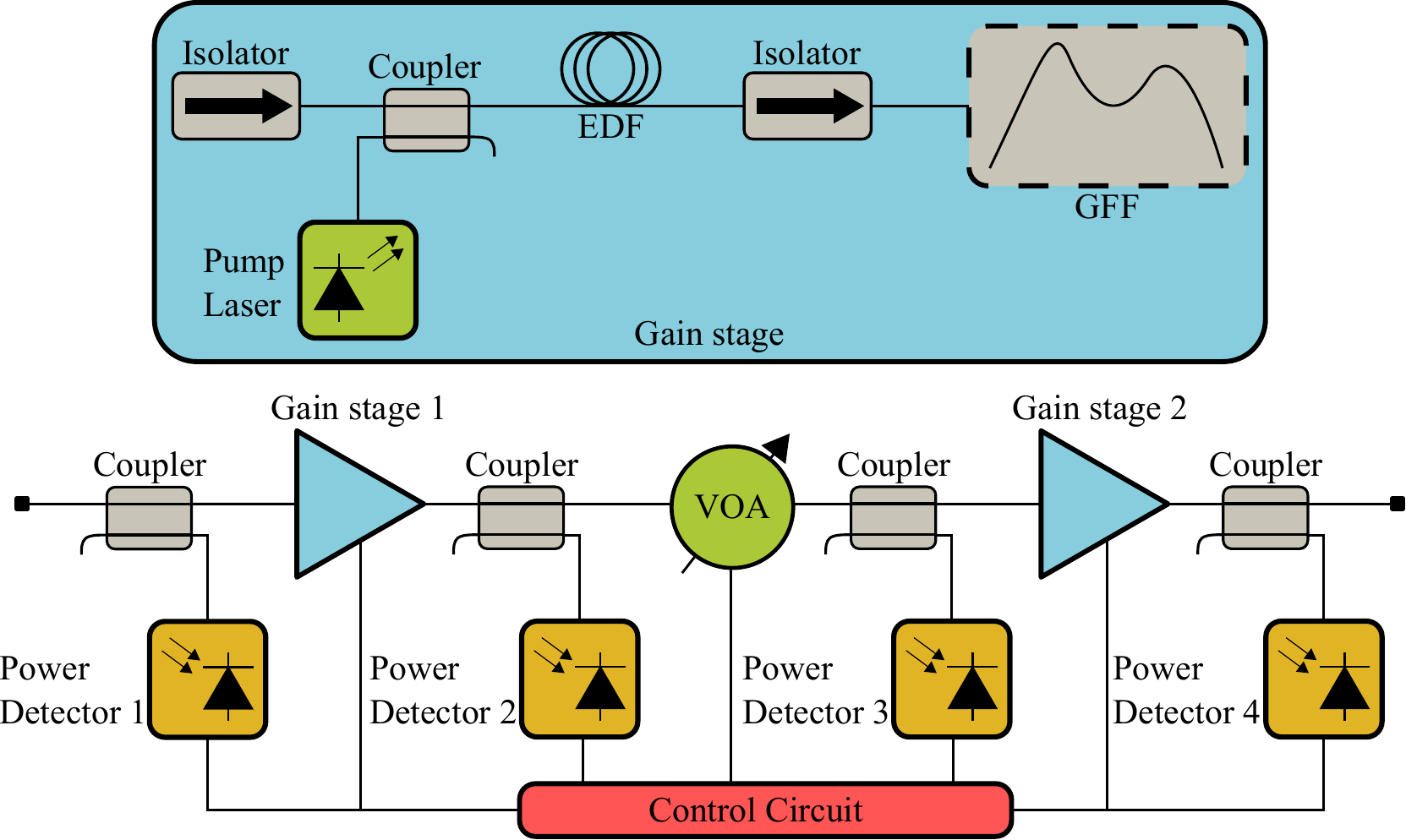}
    \caption{System architecture of the C-band EDFA}
    \label{fig:2}
\end{figure}

The diagnosis of FCs in optical fiber amplifiers is exemplified using a C-band double-stage EDFA. The architecture of the EDFA is shown in Fig.~\ref{fig:2}.

The EDFA consists of several components from a control engineering perspective, whose degradation will affect the system behavior. First, the pump lasers represent the actuators in the control loop, which excite the erbium ions in the erbium-doped fiber to amplify the optical signal. Second, the power detectors represent the sensors in the control loop, which provide feedback of the total input and output power of a stage to the controller. Third, the variable optical attenuator is another actuator, used to control the overall tilt of the EDFA system. Last, the passive components represent the components that are not actively controlled, such as the erbium-doped fiber, the isolators and the couplers. The degradation of these components can only be measured as composition of aging of all passive components, located before a power detector, resulting in four groups of passive components. Each explained case of aging will lead to a specific FC, which will be diagnosed by the model. The FCs are listed in Table~\ref{tab:1}.
\begin{table}[t]
    \centering \caption{Class mapping of the diagnosis dataset for EDFAs}
    \begin{tabular}{c l}
        \hline\hline
        \textbf{Class label} & \textbf{Normal state/ Fault case}\\
        \hline
        0 & Normal\\
        1 & Pump laser 1\\
        2 & Pump laser 2\\
        3 & Power detector 1\\
        4 & Power detector 2\\
        5 & Power detector 3\\
        6 & Power detector 4\\
        7 & Variable optical attenuator\\
        8 & Passive components 1\\
        9 & Passive components 2\\
        10 & Passive components 3\\
        11 & Passive components 4\\
        \hline\hline
    \end{tabular}
    \label{tab:1}
\end{table}

\begin{figure}[b]
    \centering
    \includegraphics[width=\linewidth]{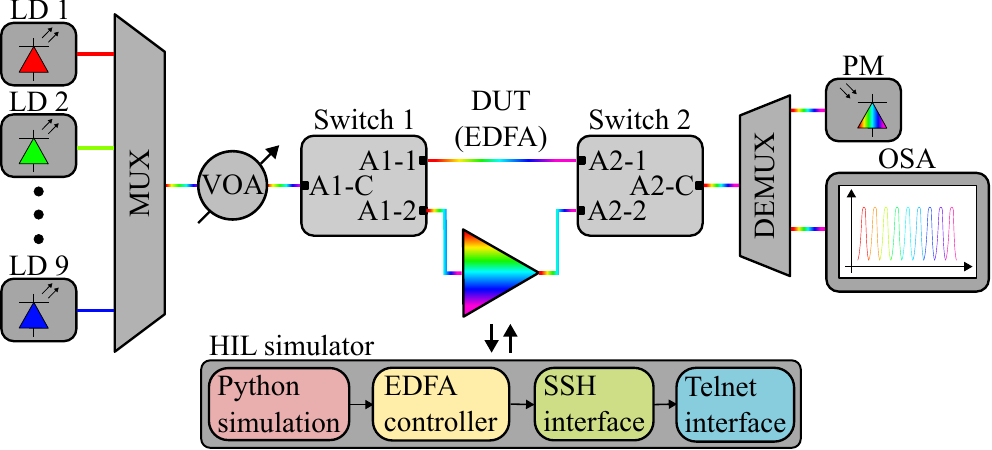}
    \caption{Data acquisition setup for the diagnosis dataset}
    \label{fig:3}
\end{figure}

The dataset is created using a hardware-in-the-loop (HIL) simulator, which generates physical-model based degradation scenarios, shown in Fig.~\ref{fig:3}. It utilizes a wavelength-division multiplexing (WDM) signal with a total of 9 channels. To vary the input power between $-35\;\mathrm{dBm}$ and $1\;\mathrm{dBm}$, a variable optical attenuator is used. The gain is varied between $19\;\mathrm{dB}$ and $35\;\mathrm{dB}$ in steps of $1\;\mathrm{dB}$. The CBM data are recorded for each operating state in a normal condition and for the HIL simulated degradation of a component. For this purpose, specific register in the internal controller of the EDFA are manipulated to simulate the degradation of a component, and record the overall system response.

The generated dataset contains a total of $11$ FCs and the normal state. The data preprocessing comprises two important mechanisms. First, the data is normalized using a standard scaler and second, a sliding time window procedure is applied \cite{9737516}. This allows the extraction of information from the dynamics of the current time point including a defined amount of previously occurred time points, which is set to $40$. Therefore, the datasets for training and testing are available as 3rd order tensors with shape $(\mathrm{Samples},\mathrm{Time Window Size},\mathrm{Features})$. The training dataset has a shape of $(2905,40,34)$ and the test dataset has a shape of $(1432,40,34)$ for each FC.

\section{Results and Discussion} 
First, the architecture of ITST is examined with regard to the influence of the three feature extraction paths in the encoder in an ablation study. The results of the ablation study are displayed in Table~\ref{tab:2}.
\begin{table}[t]
    \centering \caption{Ablation study of ITST}
    \begin{tabular}{cccc}
        \hline
        \textbf{Time Domain} & \textbf{Sensor Domain} & \textbf{Frequency Domain} & \textbf{CC}\\
        \hline
        $\checkmark$ & $\times$ & $\times$ & 0.54\\
        $\times$ & $\checkmark$ & $\times$ & 0.57\\
        $\times$ & $\times$ & $\checkmark$ & 0.61\\
        $\checkmark$ & $\checkmark$ & $\times$ & 0.43\\
        $\checkmark$ & $\times$ & $\checkmark$ & 0.47\\
        $\times$ & $\checkmark$ & $\checkmark$ & 0.47\\
        $\checkmark$ & $\checkmark$ & $\checkmark$ & 0.37\\
        \hline
    \end{tabular}
    \label{tab:2}
\end{table}

The results show that the feature extraction paths in the encoder have a positive influence on the overall classification accuracy. The best classification result with $\mathrm{CC}=0.37$ is achieved when all three feature extraction paths are used. The feature extraction in the time domain has the lowest influence on the classification result, while the feature extraction in the frequency domain has the highest influence. The feature extraction in the sensor domain has a medium influence on the classification result. The results indicate that the feature extraction paths in the encoder are complementary and improve the classification result when used together.

Second, the classification results of ITST are compared to state-of-the-art models, as shown in Table ~\ref{tab:3}. The training processes is repeated $20$ times for the purpose of statistical averaging. Here, it can be seen that ITST produces better classification results on average than the other models with $\mathrm{CC}_{mean}=0.39$. The best value for a single run, with $\mathrm{CC}_{best}=0.37$, is also better than the best values of the state-of-the-art models. Interestingly, the model with the worst classification result of $\mathrm{CC}_{mean}=0.64$ has the lowest variance with $\mathrm{CC}_{var}=1*10^{-5}$. Nevertheless, the variance of the results of the other models also remains within a small range from ${CC}_{var}=1*10^{-4}$ to ${CC}_{var}=4*10^{-4}$.
\begin{table}[b]
    \centering \caption{Classification results for diagnosing FCs in EDFAs}
    \begin{tabular}{l c c c c}
        \hline\hline
        Metrics & CNN \cite{zhao2020intelligent} & CBLSTM \cite{zhao2017learning} & Diagnosisformer \cite{hou2023diagnosisformer} & ITST\\
        \hline
        $\textrm{CC}_{mean}$ & 0.51 & 0.64 & 0.45 & \textbf{0.39}\\
        $\textrm{CC}_{var}$ & $4\cdot10^{-4}$ & $\mathbf{4\cdot10^{-5}}$ & $3\cdot10^{-4}$ & $1\cdot10^{-4}$\\
        $\textrm{CC}_{best}$ & 0.49 & 0.61 & 0.42 & \textbf{0.37}\\
        \hline\hline
    \end{tabular}
    \label{tab:3}
\end{table}
\begin{figure*}[t!]
    \centering
    \includegraphics[width=0.78\linewidth]{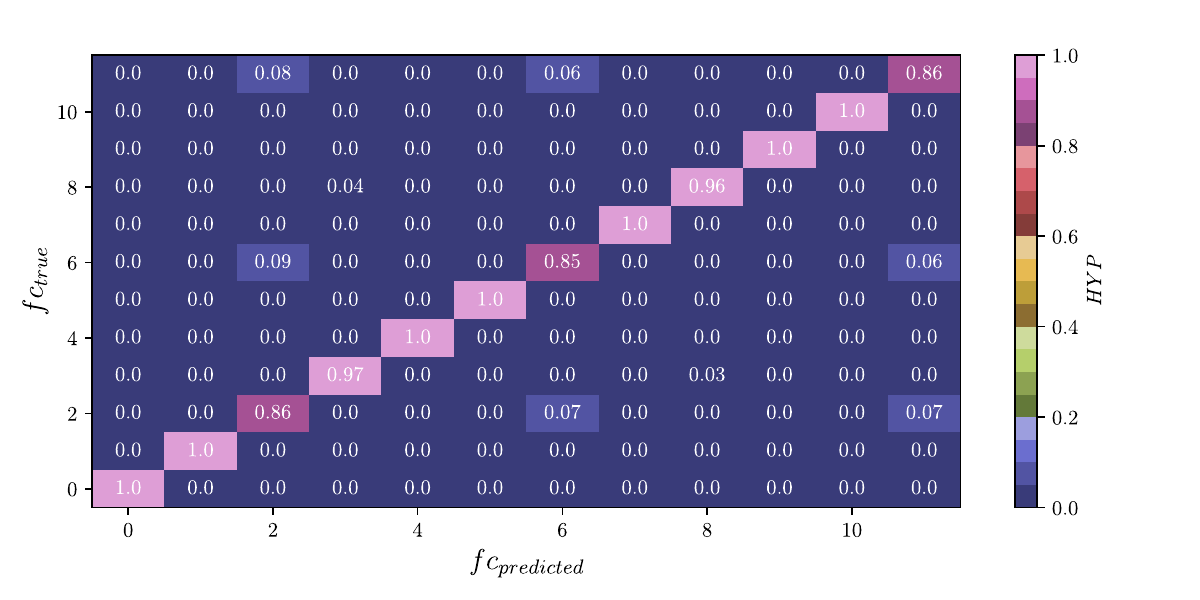}
    \caption{Confusion matrix for diagnosing FCs in EDFAs}
    \label{fig:4}
\end{figure*}
Last, the classification results of ITST are analyzed in detail. The confusion matrix for the best run of ITST is shown in Fig.~\ref{fig:4}. In this plot, the y-axis represents the actual label $fc_{true}$ of the class and the x-axis the predicted label $fc_{predicted}$. The hypothesis probability (HYP) defines the correct classification rate of the model for $fc_{true}=fc_{predicted}$ and the misclassification rate if $fc_{true}\neq fc_{predicted}$. The confusion matrix shows that the model can diagnose label 0, the normal operating state, reliably. Label 1, the degradation of the pump laser 1, is also diagnosed well. The model shows difficulties to distinguish between labels 2, 6, and 11, which represent the FC of pump laser 2, power detector 4 and passive component 4. This occurs because these FCs exhibit a high degree of similarity in their impact on the system, consequently leading to analogous data patterns. In addition, there is a slight tendency to confuse label 2 with label 8, representing the FC of power detector 1 and passive component 1. All other labels, meaning FCs, are diagnosed very well by ITST.

In summary, it can be said that ITST can reliably identify the individual occurrence of various FCs based on the CBM time series data. However, it should be noted that only investigations regarding the occurrence of one FC at a specific point in time are carried out. There are two reasons for this. First, considering a combination of occurrences of multiple FCs simultaneously transforms a problem of order $\mathcal{O}(N_{FC})$, with $N_{FC}=12$ as the number of FCs, into a problem of order $\mathcal{O}(2^{N_{FC}})$. Secondly, it is unlikely that two FCs occur simultaneously, as these degradation processes are caused by a particular component in the system. The observation of an isolated event is more likely with the long degradation times.

\section{Conclusion} 
In this work, a novel transformer-based model, Inverse Triple-Aspect Self-Attention Transformer (ITST), is presented, which enables the diagnosis of FCs in optical fiber amplifiers using CBM time series data. The model uses an encoder-decoder architecture with three feature extraction paths in the encoder, feature-engineered data in the decoder and the self-attention mechanism. An ablation study has been carried out to demonstrate the effectiveness of the proposed architecture. The classification results show that ITST outperforms state-of-the-art models in terms of classification accuracy. The diagnosis of FCs in optical fiber amplifiers is exemplified using a C-band EDFA for the first time. This enables the early detection of degraded components, allowing maintenance measures to be planned and network downtimes to be reduced. Future work will focus on the application of ITST to other optical fiber amplifier types. Furthermore, the model will be extended to include the diagnosis of FCs in other optical network components.

\bibliographystyle{IEEEtran}
\bibliography{IEEEabrv, references.bib}

\end{document}